\documentclass[english,RNAAS,tighten]{aastex631}
\usepackage[T1]{fontenc}
\usepackage[latin9]{inputenc}
\usepackage{graphicx}

\makeatletter
\shorttitle{How much is one bit?}
\shortauthors{Johannes Buchner}

\makeatother

\usepackage{babel}
\begin{document}
\title{AGN sub-populations important for black hole mass growth: a rule of thumb}
\author[0000-0003-0426-6634]{Johannes Buchner} \affiliation{Max Planck Institute for Extraterrestrial Physics, Giessenbachstrasse, 85741 Garching, Germany}

\begin{abstract}For building the super-massive black hole population within a Hubble
time, only sub-populations with more than $10^{52}\mathrm{erg/s}/\overline{L}$
objects on the sky are relevant, where $\overline{L}$ is the sample-averaged
bolometric luminosity.

\end{abstract}

\section{Building mass}

To build a super-massive black hole with $M_{\bullet}=10^{8}\,M_{\odot}$
within the age of the Universe $\Delta t$ requires an average accretion
luminosity of $\overline{L}_{\mathrm{bol}}=5\times10^{43}\,\mathrm{erg/s}$,
assuming the usual luminosity efficiency conversions ($L_{\mathrm{bol}}=\frac{\epsilon}{1-\epsilon}\dot{M}c{{}^2}$,
$\epsilon=10\%$). This is slightly above the common threshold for
what is considered an active galactic nucleus (AGN), $L_{\mathrm{X}}>10^{42}\,\mathrm{erg/s}$,
with the bolometric conversion $L_{\mathrm{X}}(2-10\mathrm{keV})\approx L_{\mathrm{bol}}/10$
\citep[e.g.,][]{Marconi2004}. Smaller black holes $M_{\bullet}=10^{4}M_{\odot}$
can be assembled with $\overline{L}_{\mathrm{bol}}=5\times10^{39}\,\mathrm{erg/s}$. 

In considering the average luminosity, it is not necessary to assume
a constant luminosity over $\Delta t$. Indeed, the accretion rate
may be fluctuating over many orders of magnitude over long time-scales
\citep[e.g.][]{Sartori2019}, and most galaxies are not AGN. This
means that if the black hole is active only in brief phases, the observed
luminosities need to be higher to reach the same average luminosity. 

\section{Building mass density}

In the local Universe, the mass density locked into black holes is
estimated to be $\rho_{\bullet}\approx5\times10^{5}\,M_{\odot}/\mathrm{Mpc}^{3}$
\citep[e.g.,][]{Marconi2004}. Which sub-populations of AGN contribute
substantially ($>10\%$) to black hole growth history? Ten per cent
($50,000\,M_{\odot}/\mathrm{Mpc}^{3}$) corresponds to an average
luminosity density over the age of the Universe $\Delta t$ of:

\begin{equation}
10\%\times\rho_{\bullet}/\Delta t\times\left(\frac{\epsilon}{1-\epsilon}c{{}^2}\right)=\overline{L\times\phi}=2\times10^{40}\,\mathrm{erg/s/Mpc}^{3}\label{eq:averagedensity}
\end{equation}

For a substantial contribution (eq. \ref{eq:averagedensity}), sources
need to be luminous but also \emph{numerous}. They need to be numerous
to enable a high on-time (duty cycle) of the accretion process, ortherwise
the luminosity averaged over cosmic time is small. The expected number
of AGN on the entire sky can be estimated by $N=V_{\mathrm{c}}\times\overline{\phi}$.
We use the comoving volume $V_{\mathrm{c}}=2\times10^{12}\,\mathrm{Mpc}^{3}$
out to $z=4$, as the number density of AGN strongly declines above
$z=3$ \citep[e.g.,][]{Georgakakis2015}. This gives the rule of thumb
that samples that contribute more than 10 percent of black hole growth
need to have at least $10^{52}\,\mathrm{erg/s}/\overline{L}$ sources
in the Universe.

\section{Examples}

Lets first consider the \emph{most extreme AGN}. For hot dust-obscured
galaxies \citep[DOGs,][]{Tsai2015}, $\overline{L}=10^{48}\mathrm{erg/s}$.
Therefore, $\overline{\phi}=10^{-8}\mathrm{Mpc}^{-3}$, which means
on the entire sky, at least $N=V_{c}\times\overline{\phi}=10^{4}$
objects need to be seen for this population to be significant. However,
only a few hundred objects were found, indicating that hot DOGs are
not an important black hole mass growth phase. This can be either
because hot DOGs are created too rarely, or because the phase is too
brief.

\begin{figure}
\begin{centering}
\includegraphics[width=0.7\textwidth]{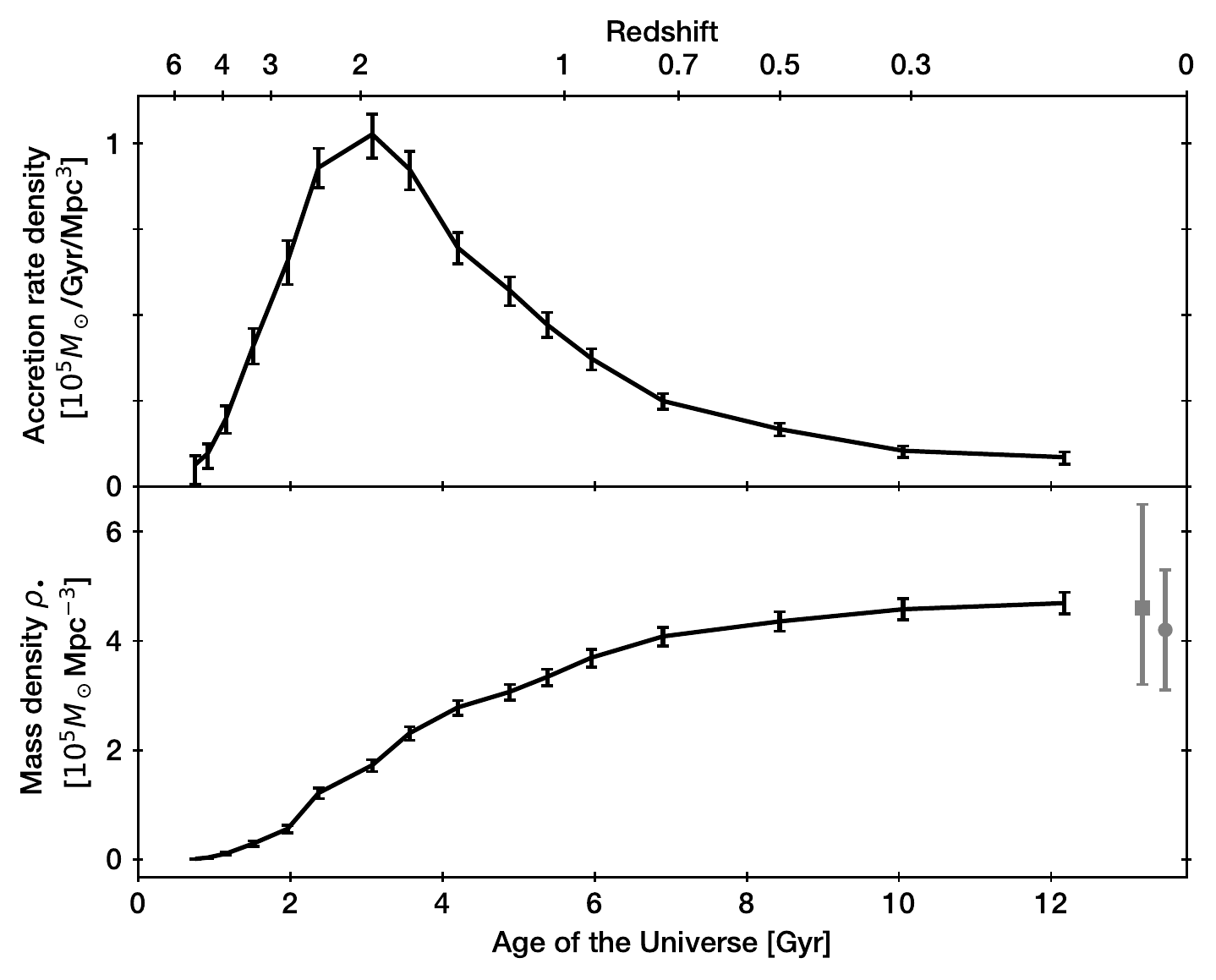}
\par\end{centering}
\caption{\label{fig:BHMD}Top panel: Black hole accretion history inferred
from X-ray luminosity functions, including Compton-thick growth. Bottom
panel: accumulated mass density over cosmic time. The gray error bars
are local densities estimates are from \citet{Marconi2004} and \citet{Shankar2004}.
Most black hole growth occurs after the first and before the last
2 Gyrs.}

\end{figure}

Considering a more moderate luminosity, $\overline{L}=10^{46}\mathrm{erg/s}$,
such as for merger-induced red quasars \citep[e.g.,][]{Urrutia2008},
then $\overline{\phi}=10^{-6}\mathrm{Mpc}^{-3}$, which means on the
entire sky, at least $N=10^{6}$ objects, or 24 objects per square
degree, are required. However, only a few hundred objects have been
found, suggesting that \emph{major merger phases} are too rare to
contribute substantially to black hole growth.

Finally, let us consider the redshift interval where substantial growth
occurs. Figure~\ref{fig:BHMD} plots the accretion history inferred
from the X-ray luminosity function of \citet[slope-constant machine-learned space density model]{Buchner2015},
and includes unobscured, mildly obscured and the most heavily obscured
AGN population. The top panel illustrates that in the \emph{local
Universe}, the current accretion rate density is an order of magnitude
lower than at its peak at cosmic noon, where it was $6\times10^{41}\mathrm{erg/s/Mpc}^{3}/\mathrm{Gyr}$
. For this reason, the growth accumulated (bottom panel) in the last
2 Gyrs (since $z<0.3$) is small and amounts to less than 5 per cent.
The growth in the first Gyr is also negligible.

\section{Closing words}

While extreme and rare sub-populations may not contribute substantially
to black hole growth, they are laboratories that expose extreme physical
processes most clearly. However, when super-massive black hole mass
assembly is considered, sub-populations with more than $10^{52}\mathrm{erg/s}/\overline{L}$
objects on the sky, and $\overline{L}>5\times10^{39}\mathrm{erg/s}$
are required.

\section*{Acknowledgements}

I thank Isabelle Gauger and Qingling Ni for helpful feedback on the
manuscript. This work was inspired by discussions in the ``What drives
the Growth of Black Holes? A decade of reflection'' workshop 2022.

\bibliographystyle{apj}
\bibliography{agn,stats}

\end{document}